\documentclass[a4paper]{jpconf}
\usepackage{graphicx}

\newcommand{\pT} {\ensuremath{p_{\mathrm{T}}}}
\newcommand{\pseudorap} {\mbox{$\left | \eta \right | $}}
\newcommand{\ppbar} {\mbox{$\mathrm{p}\kern-0.05em \mathrm{\overline{p}}$}}
\newcommand{\pandp} {\mbox{$\mathrm{p}\kern-0.05em \mathrm{p}$}}

\newcommand {\tev} {\mbox{${\rm TeV}$}}

\newcommand {\mom} {\mbox{\rm GeV$\kern-0.15em /\kern-0.12em c$}}
\newcommand {\mmass} {\mbox{\rm MeV$\kern-0.15em /\kern-0.12em c^2$}}

\newcommand {\cm} {\mbox{${\rm cm}$}}

\newcommand{\rmLambda}{\mbox{$\mathrm {\Lambda^{0}}$}}
\newcommand{\rmAlambda}{\mbox{$\mathrm {\overline{\Lambda}^{0}}$}}
\newcommand{\rmXi}{\mbox{$\mathrm {\Xi^{-}}$}}
\newcommand{\rmAxi}{\mbox{$\mathrm {\overline{\Xi}^{+}}$}}
\newcommand{\rmOmega}{\mbox{$\mathrm {\Omega^{-}}$}}
\newcommand{\rmAomega}{\mbox{$\mathrm {\overline{\Omega}^{+}}$}}
\newcommand{\Xis}{\mbox{$\mathrm {\Xi^{-}+\overline{\Xi}^{+}}$}}

\usepackage{cite}
\begin{document}
\title{Multi-strange baryon measurements at LHC energies, with the ALICE experiment}

\author{Antonin Maire for the ALICE Collaboration}

\address{Institut pluridisciplinaire Hubert Curien (IPHC), 23 rue du loess - 67037 Strasbourg, France }

\ead{antonin.maire@cern.ch}

\begin{abstract}
The status of the charged multi-strange baryon analysis (\rmXi, \rmAxi, \rmOmega, \rmAomega) at LHC energies is presented. This report is based on the results obtained with ALICE (A Large Ion Collider Experiment), profiting from the characteristic cascade-decay topology. A special attention is drawn to the early \pandp~data-taking period (2009-2010) and subsequently, on the uncorrected \pT-spectra extracted at mid-rapidity for centre of mass energies of 0.9~\tev~and 7~\tev.
\end{abstract}

%
\section{Introduction and motivation}
\label{sec:Intro}

Measuring strange particle production in proton-proton (\pandp) collisions is a necessary benchmark for the physics of ultra relativistic heavy ions. This is important at the Large Hadron Collider (LHC), where the heavy-ion programme is scheduled to begin in November 2010~\cite{Alessandro:2006yt}. 
Moreover, strangeness in \pandp~collisions is interesting in itself, as it may shed light on hadron production mechanisms.
While the \emph{hard} component of the event may be described by the perturbative Quantum Chromo-Dynamics (based on parton-parton scattering and fragmentation~\cite{Sjostrand:2006za,Engel:1995sb}), the \emph{soft} component must be treated in a more complex manner.
Currently, the soft physics is described via thermal models~\cite{Becattini:1997rv, BraunMunzinger:2003zd} or via QCD-inspired models (relying on multi-parton processes~\cite{Sjostrand:2004ef} or multiple scattering~\cite{Werner:2008zza, Werner:2010aa}, for instance). In either case, further improvements of such phenomenological approaches may be spurred by confrontation with experimental measurements.

In that respect, double- and triple-strange baryons ($\Xi$, $\Omega$, ...), which are the focus of this publication, may provide relevant insights: due to identification via weak-decay topology reconstruction, they can be studied over a large momentum range. Starting from \pT~$\approx 0.6~\mom$ and up to $\approx 10~\mom$, these spectra cover the region dominated by the soft processes and reach the energy scale where hard scattering mechanisms dominate. 

Some measurements have already been performed at previous and current facilities. These include both \ppbar~collisions (S$\mathrm{p\overline{p}}$S, Tevatron) and \pandp~collisions (RHIC), with centre of mass energies $\sqrt{s}$ ranging from 0.2~\tev~up to 1.96 \tev~\cite{Abelev:2006cs,Ansorge:1989ba,CDFrunII:2010h}. The LHC, in operation since November 2009, can extend the existent 0.9-\tev~measurements made by the UA5 collaborations in \ppbar, and perform new measurements at $\sqrt{s}$ = 7~\tev, beyond the Tevatron energies.

The ALICE experiment~\cite{Alessandro:2006yt} is well-suited for such spectrum measurements, due to a low \pT~cut-off and excellent particle identification (PID) capabilities. The low \pT~cut-off is made possible by the low magnetic field applied in the central barrel ($\leq 0.5~T$) and the low material budget in this mid-rapidity region ($13\%$ of radiation length~\cite{Hippolyte:2009xz}). The PID capabilities are supplied by a set of detectors utilizing diverse techniques (energy loss, time of flight, transition radiation, Cerenkov effect).

%
\section{Data analysis and identification methods}
\label{sec:DaId}

\subsection{Data collection and detector setup}
\label{ssec:DetLHC}

The data presented here is from the minimum bias sample collected during the Nov-Dec 2009 LHC \pandp~run at $\sqrt{s}$ = 0.9~\tev~\cite{Evans:2008zzb} ($\sim3 \times 10^{5}$ events), and from the 7 \tev~\pandp~run that started in March 2010 and is ongoing ($\sim23.9 \times 10^{6}$ events were analysed here but more than $4 \times 10^{8}$ events have been collected since then).
This study makes use of the ALICE central barrel~\cite{Aamodt:2008zz}, covering a pseudo-rapidity range \pseudorap $< 0.9$ and the full azimuth, the whole being embedded in the large L3 solenoidal magnet which provides a nominal magnetic field of $0.5~T$. 

The cascade signals are obtained using essentially data collected by the two main tracking detectors: the Inner Tracking System (ITS), composed of 6 cylindrical layers of high-resolution silicon detectors~\cite{Aamodt:2010its}, and the cylindrical Time Projection Chamber (TPC)~\cite{Alme:2010tpc}.

\subsection{Topological reconstruction}
\label{ssec:TopoReco}

The multi-strange hadron identification is performed using a combination of displaced-vertex reconstruction, invariant mass analyses as well as single track PID methods, such as energy loss in the TPC.
The reconstruction of the different multi-strange particles hinges on their respective charged weak decays, the so-called \emph{cascade} structures.
For each particle of interest, the main characteristics and utilized decay channels are listed in table \ref{Tab:maincharac}. The anti-baryons are similarly reconstructed via the decay channel involving the charge conjugates. 

\begin{table}[!hbt]
\begin{center}
    \caption{Main characteristics of the reconstructed particles~\cite{Amsler:2008pdg}.}
    \label{Tab:maincharac}
    \begin{tabular}{llcclc}
      \br
         Particles               &   mass ($\mmass$)   &  c$\tau$   & charged decay & B.R. \\
      \mr
        $\rmLambda$ ($uds$)              & $1115.68$   & $7.89~\cm$ &  $\rmLambda \rightarrow p + \pi^-$             & 63.9\% \\
        $\rmXi$  ($dss$)                 & $1321.71$   & $4.91~\cm$ &  $\rmXi \rightarrow  \rmLambda+ \pi^-$         & 99.9\% \\
        $\rmOmega$ ($sss$)               & $1672.45 $  & $2.46~\cm$ &  $\rmOmega \rightarrow  \rmLambda+ \rm{K}^-$   & 67.8\% \\
      \br
    \end{tabular}
\end{center}
\end{table}

\begin{figure}[!hbt]
      \includegraphics[width=0.99\textwidth]{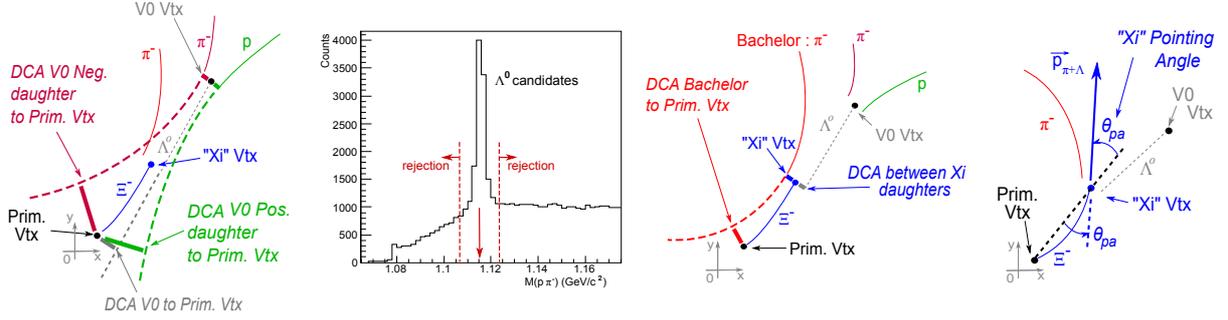}
      \caption{Cascade (\rmXi, \rmAxi, \rmOmega~and \rmAomega) reconstruction principle. The acronym \emph{DCA} stands for Distance of Closest Approach.}
      \label{Fig:CascReco}
\end{figure}

The guidelines of the reconstruction algorithm dedicated to cascades are sketched in Fig. \ref{Fig:CascReco}. The \rmXi, \rmAxi, \rmOmega~and \rmAomega~identification is based on the pairing of a \rmLambda~or \rmAlambda~baryon with an additional particle called \emph{bachelor}.
The \mbox{$\mathrm {\Lambda}$} reconstruction is grounded in the secondary vertex finding, a \emph{V0} structure built out of two \emph{secondary} tracks of opposite charges, compatible with coming from the same vertex. It is checked that the V0 candidate thus obtained sits in the proper invariant mass window to validate the \mbox{$\mathrm {\Lambda}$} hypothesis. 
The \mbox{$\mathrm {\Lambda}$} candidate is then matched with a \emph{secondary} track, to form a typical cascade structure. The reconstructed momentum of the multi-strange baryon must point in the direction of the primary vertex.
The V0 vertex and the V0-bachelor matching are limited to a certain fiducial volume. 
A PID compatibility cut, based on the dE/dx measurement from the TPC, is required for each of the three decay products, known as \emph{daughter} particles, to partially remove some combinatorial background. 

\subsection{Signal extraction}
\label{ssec:SigExtract}

\begin{figure}[!hbt]
   \begin{minipage}[c]{.46\linewidth}
      \includegraphics[width=0.99\textwidth]{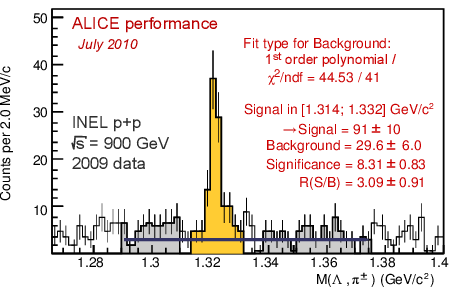}
      \caption{Signal extraction based on ``bin counting'' method, illustrated with the \Xis invariant mass distribution in $1.4< \pT~(\mom)<2.0$.}
      \label{Fig:SignalExtraction}
   \end{minipage} \hfill
   \begin{minipage}[c]{.46\linewidth}
 \includegraphics[width=0.99\textwidth]{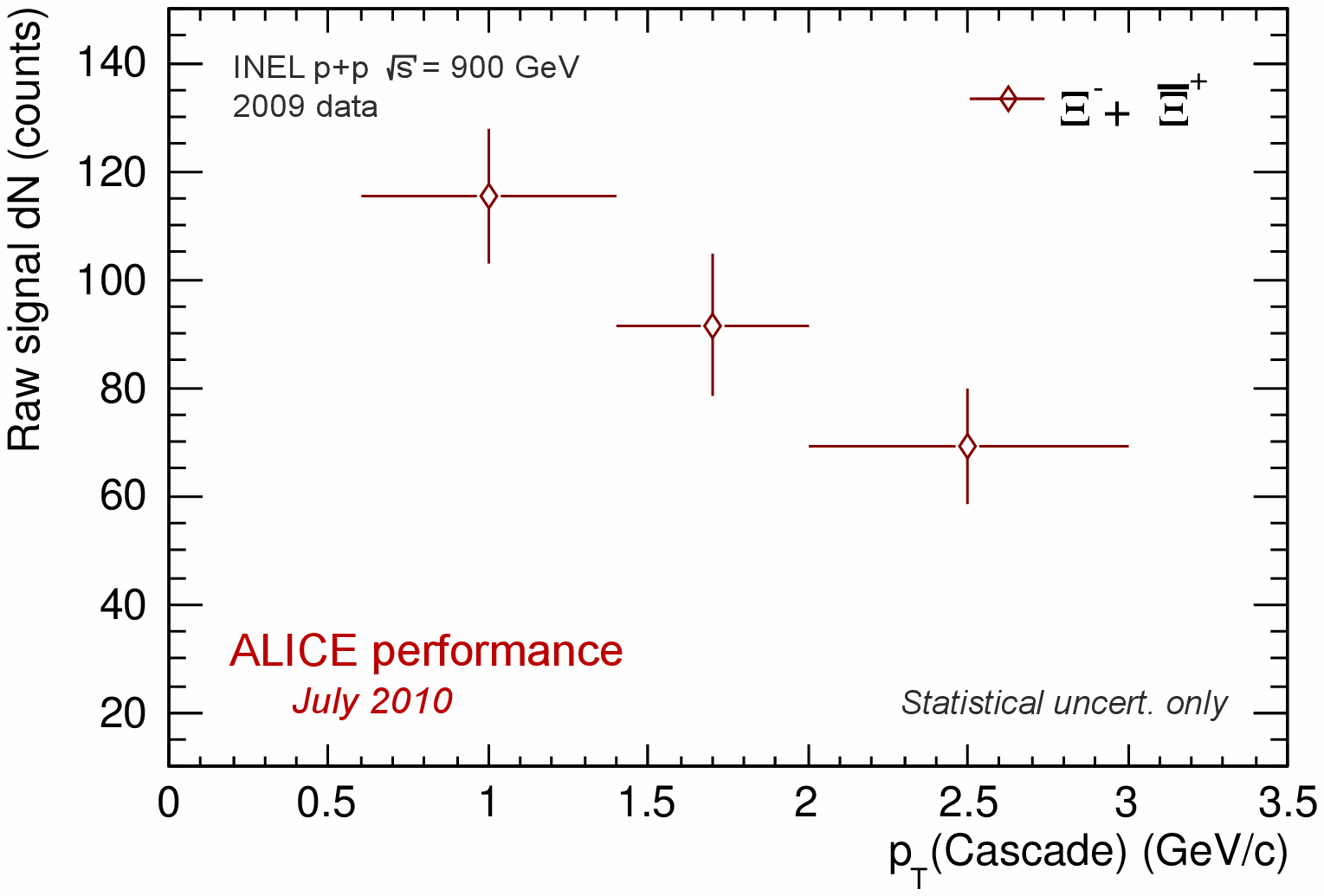}
      \caption{Raw signal by \pT~bins for \Xis, in 2009 \pandp~data at 0.9~\tev.}
      \label{Fig:RawYieldsXis900}       
    \end{minipage} \hfill
\end{figure}

For each considered particle, we intend to extract a signal in successive \pT~intervals (bins). For each \pT~bin, the invariant mass signal sits on top of a given background. The signal extraction process, using ``bin-counting'' method, is illustrated in Fig. \ref{Fig:SignalExtraction}. 

The signal is first approximated by a Gaussian lying on a polynomial background. Because of the non-Gaussian tails of the signal, this results in rough but sufficient estimates of the signal mean and width.

The background on each side of the signal (more than $6\sigma$ away from the Gaussian mean) is then sampled. 
The width of each background area is taken to be $7\sigma$ large.

The sum of signal and background ($S$+$B$) is sampled in the region defined by the Gaussian mean $\pm 4\sigma$. Consequently, we make use of the areas previously sampled on the side-signal bands to assess the background $B$ under the signal $S$. The signal yield $S=(S+B)-B$ is thus computed without any assumption as to the signal shape.

%
\section{Measurements at 0.9 and 7~\tev}
\label{sec:Meas}

The results for the 2009 \pandp~sample at 0.9~\tev~are shown in Fig. \ref{Fig:RawYieldsXis900}. The plot shows the signal counts (raw yields) for \Xis, as a function of \pT. The uncertainties correspond to both the statistical uncertainty related to the number of counts and the uncertainty from signal extraction. 

%

\begin{figure}[!hbt]
    \begin{minipage}[c]{.46\linewidth}
      \includegraphics[width=0.95\textwidth]{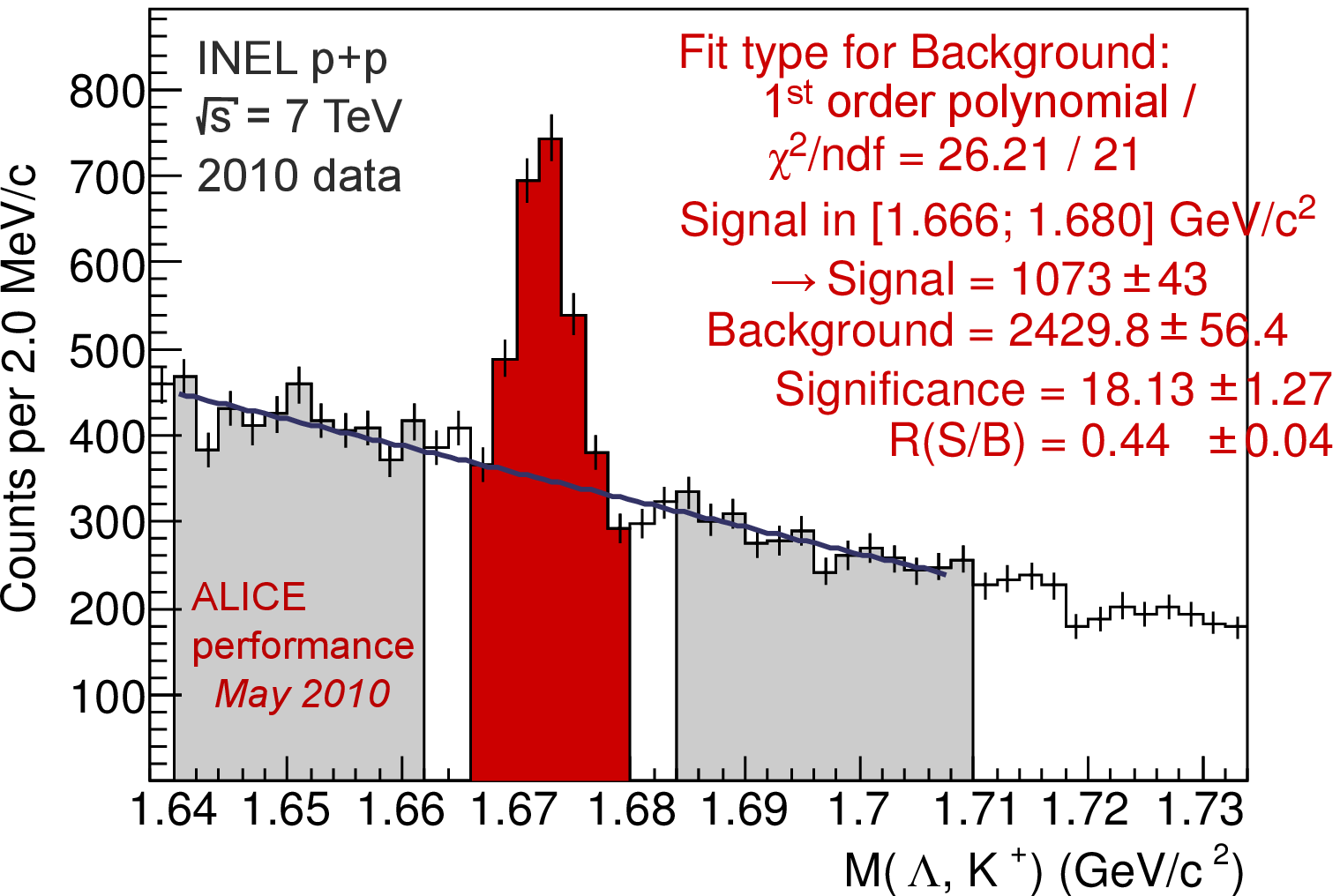}
      \caption{Invariant mass showing the \rmAomega~signal in 2010 \pandp~data at 7 \tev~(23.9~M events).}
      \label{Fig:OmegaMinusInvMass}
   \end{minipage} \hfill
   \begin{minipage}[c]{.46\linewidth}
        \includegraphics[width=0.99\textwidth]{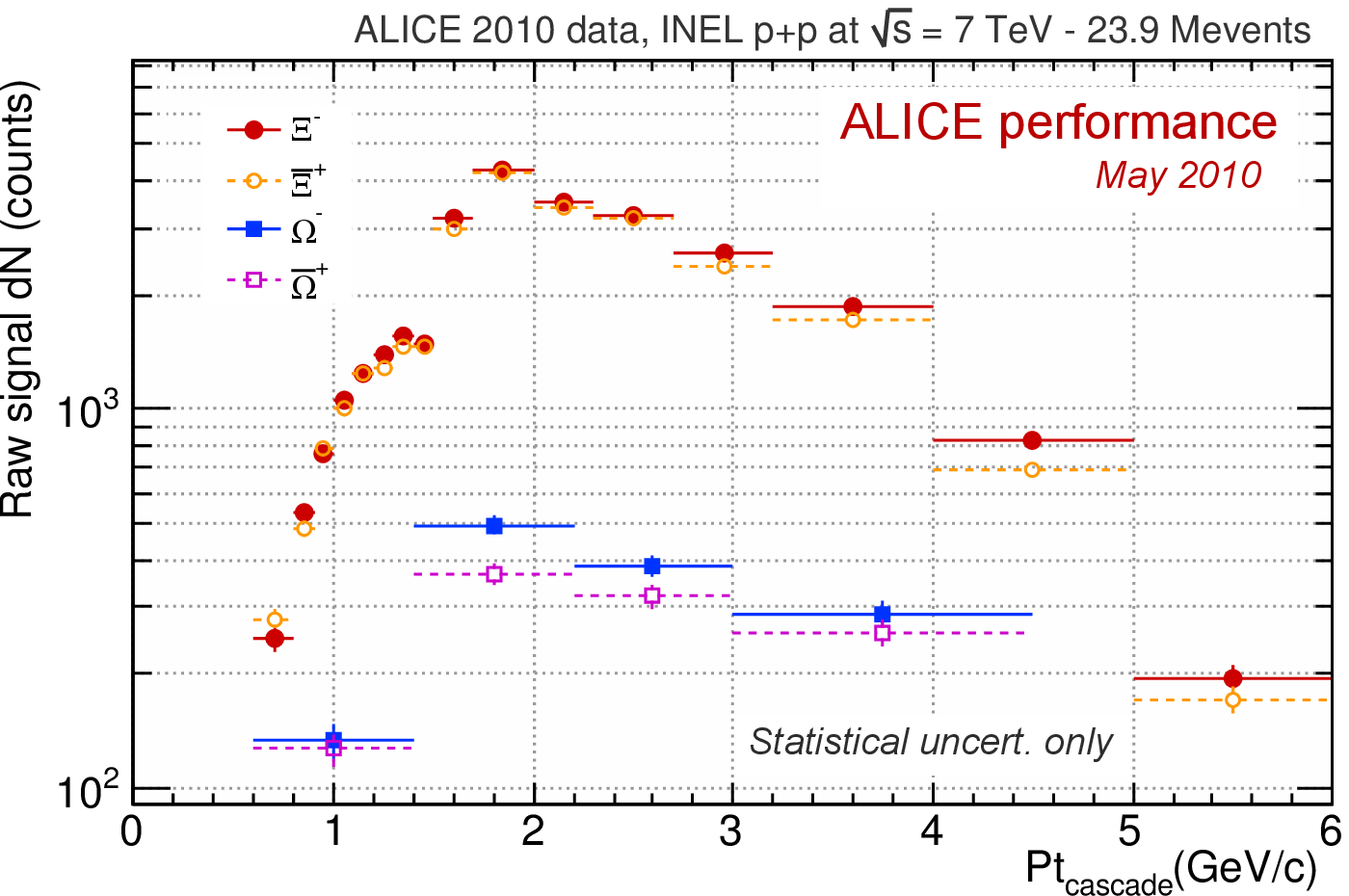}
        \caption{Raw signal by \pT~bins, extracted for the different cascade species, in 2010 \pandp~data at 7 \tev~(23.9~M events).}
        \label{Fig:UncorrSpectra7TeV}
   \end{minipage}
\end{figure}

 The same particles are presently analysed for the \pandp~data at 7 \tev. Due to higher statistics, baryons and anti-baryons can be studied separately and with larger counts: \rmXi~and \rmAxi~or even \rmOmega~and \rmAomega~can now be distinguished from one another, as illustrated in Fig.~\ref{Fig:OmegaMinusInvMass} and \ref{Fig:UncorrSpectra7TeV}.

%
\section{Summary}
\label{sec:Ccl}

A spectrum for \Xis~baryons is obtained for the first LHC \pandp~run at 0.9~\tev, despite the limited statistics; the assessment of efficiency corrections and systematic uncertainties is under way.
Due to the large statistics available for the 7-\tev~\pandp~data sample, \rmXi, \rmAxi, \rmOmega~and \rmAomega~can be extracted distinctively.
This bodes well for more accurate and differential analyses such as spectra as a function of \pT, rapidity or event multiplicity.


\end{document}